\documentclass[10pt]{amsart}
\usepackage{amssymb}
\usepackage{enumerate}
\usepackage{epsf}
\usepackage{psfrag}
\DeclareGraphicsExtensions{.eps,.art,.ART,.ps}
\newcommand{\TA}{T\!A}

\newcommand{\bbR}{\mathbb{R}}      
\newcommand{\bbN}{\mathbb{N}}      

\newcommand{\arccosh}{\operatorname{arccosh}}
\newcommand{\arctanh}{\operatorname{arctanh}}
\newcommand{\cotanh}{\operatorname{cotanh}}
\begin{document}
\begin{abstract}
Using the theory of optimal rocket trajectories in general relativity, recently developed in \cite{HN11}, we present a candidate for the minimum total integrated acceleration closed timelike curve in the G\"odel universe, and give evidence for its minimality. The total integrated acceleration of this curve is lower than Malament's conjectured value \cite{Malament84}, as was already implicit in the work of Manchak \cite{Manchak11}; however, Malament's conjecture does seem to hold for periodic closed timelike curves.
\end{abstract}
%
%
\title{Optimal time travel in the G\"odel universe}
\author{Jos\'{e} Nat\'{a}rio}
\address{Centro de An\'alise Matem\'atica, Geometria e Sistemas Din\^amicos, Departamento de Matem\'atica, Instituto Superior T\'ecnico, 1049-001 Lisboa, Portugal}
\thanks{Partially supported by FCT (Portugal).}
\maketitle
%
%
%
\section*{Introduction}
It is well known that the G\"odel universe contains closed timelike curves \cite{Godel, HE95}. These curves are not geodesics, and so any particle following them must accelerate. G\"odel himself remarked that if the acceleration is provided by a rocket then the required amount of fuel is prohibitively high \cite{Godel49, Malament87, OS03}. This amount can be computed from the well known rocket equation \cite{Ackeret46, HN11}, which relates the initial and final rest masses of the rocket:
\[
m(\tau_1) = m(\tau_0) \exp\left(-\frac1v \int_{\tau_0}^{\tau_1} a(\tau) d\tau\right),
\]
where $\tau$ is the proper time, $m(\tau)$ is the rest mass, $a(\tau)$ is the magnitude of the proper acceleration and $v$ is the ejection speed. Since $v$ must be smaller than the speed of light\footnote{We use the conventions of \cite{MTW73}, including geometrized units ($c=G=1$).}, $v \leq 1$, we have
\[
\frac{m(\tau_0)}{m(\tau_1)} \geq \exp(\TA),
\]
where
\[
\TA=\int_{\tau_0}^{\tau_1} a(\tau) d\tau
\]
is the so-called {\bf total integrated acceleration}. Malament \cite{Malament84, Malament85} proved that $\TA \geq \ln(2 + \sqrt{5}) \simeq 1.4436$ for any closed timelike curve in the G\"odel universe, and conjectured that in fact $\TA \geq 2\pi\sqrt{9+6\sqrt{3}} \simeq 27.6691$. Notice that if this conjecture is true then
\[
\frac{m(\tau_0)}{m(\tau_1)} \gtrsim 10^{12}
\]
for any rocket following a closed timelike curve in the G\"odel universe! Manchak \cite{Manchak11}, however, proved that there exist closed timelike curves in the G\"odel universe with $\TA < 2\pi\sqrt{9+6\sqrt{3}}$. These curves are not periodic, in the sense that the initial four-velocity is not equal to the final four-velocity, and it seems likely that if one includes the integrated acceleration necessary to boost the final four-velocity to the initial four-velocity then Malament's conjecture still holds. Nevertheless, Manchak did succeed in showing that one can have time travel in the G\"{o}del universe with total integrated acceleration smaller than Malament's bound.

\begin{figure}[h!]
\begin{center}
\psfrag{1/n}{$\frac1n$}
\epsfxsize=.7\textwidth
\leavevmode
\epsfbox{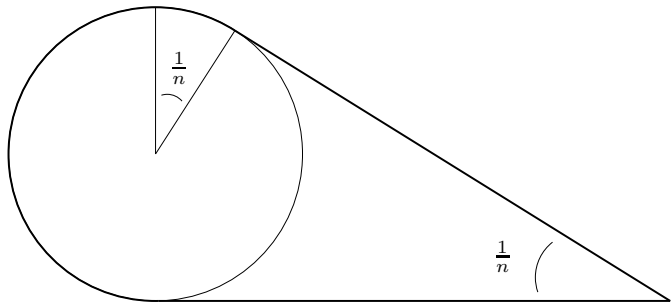}
\end{center}
\caption{Sequence of closed curves with total integrated geodesic curvature approaching $\pi$.}\label{Closed_Euclidean}
\end{figure}

The situation is analogous to the problem of finding a closed curve in Euclidean space for which the total integrated geodesic curvature with respect to the arclength (the Riemannian analogue of the Lorentzian total integrated acceleration) is minimum. If we require the curve to be periodic, then the minimum is $2\pi$, and it is attained by a circle, for example. If, however, we do not require the initial and final directions of the curve to match up, there is no minimum; the infimum is $\pi$, and is approached by a sequence of closed curves consisting of an arc of circle subtending an angle $\pi + \frac1n$ (for $n \in \bbN$) and two line segments tangent to its endpoints, which therefore meet at an angle $\frac1n$ (see Figure~\ref{Closed_Euclidean}). The intersection of the line segments is meant to be the curve's initial and final point. Notice that the integrated geodesic curvature required for ``turning the corner'' at this point (thus making the curve periodic) is precisely $\pi-\frac1n$.

In this paper we present a candidate for the minimum total integrated acceleration closed timelike curve in the G\"odel universe, and give evidence for its minimality. The total integrated acceleration of this curve turns out to be approximately $24.9927$. If one includes the integrated acceleration necessary to boost the final four-velocity to the initial four-velocity, however, total integrated acceleration increases to about $28.6085$, and so Malament's conjecture is not violated.

The candidate curve is more complicated than one would guess from the Euclidean space analogy, and is obtained by applying the general theory of optimal rocket trajectories in general relativity developed in \cite{HN11}. For the reader's convenience, this theory is summarized in section~\ref{section1}, and the main features of the G\"{o}del universe are reviewed in section~\ref{section2}. In section~\ref{section3} we solve the Jacobi equation for a timelike geodesic, and in section~\ref{section4} we compute the circular optimal accelerated arcs, both of which are used in section~\ref{section5} to construct the candidate curve and give evidence for its minimality.
%
%
\section{The rocket problem in general relativity}\label{section1}
In \cite{HN11} the classical Newtonian theory of optimal rocket trajectories developed in \cite{Lawden63} was generalized to the general relativity setting. There it is shown that optimal trajectories are expected to be continuous, sectionally smooth timelike curves, obtained by piecing together free-fall (geodesic) and accelerated arcs, possibly with instantaneous (Dirac delta) accelerations at the junction points. If the trajectory minimizes the total integrated acceleration then it satisfies the differential equations
\begin{equation} \label{diffeqns}
\begin{cases}
\displaystyle \nabla_U U^\mu = a P^\mu \\
\nabla_U P^\mu = - q^\mu + a U^\mu \\
\nabla_U q_\mu = R_{\mu\alpha\beta\gamma}U^\alpha P^\beta U^\gamma
\end{cases}
\end{equation}
where $U$ is the four-velocity, $a$ is the magnitude of the proper acceleration, $P$ is an auxiliary vector field, called the {\bf primer}, and $R_{\alpha\beta\mu\nu}$ are the components of the Riemann curvature tensor. All quantities above are continuous at a junction point except possibly $U$ and $P$, which change by a boost with positive parameter\footnote{This boost parameter is exactly the contribution of the instantaneous acceleration to the total integrated acceleration} in their common plane if there is an instantaneous acceleration. The magnitude of $\rho$ the primer, however, is $C^2$ at the junction points, and satisfies $\rho \leq 1$ on free-fall arcs and $\rho=1$ on accelerated arcs and instantaneous accelerations. Moreover, the equations above admit the first integrals 
\begin{equation} \label{firstintegrals}
P_\mu U^\mu = q_\mu U^\mu = 0.
\end{equation}
Finally, it is easily seen from \cite{HN11} that if the initial or final four-velocities are not specified then we must have $P=0$ at those points.

For free-fall arcs ($a=0$), conditions \eqref{diffeqns} reduce to the geodesic and Jacobi (geodesic deviation) equations, with the first integrals \eqref{firstintegrals} requiring the Jacobi field $P$ to be orthogonal to the four-velocity $U$. 

For accelerated arcs, we have from \eqref{diffeqns} and \eqref{firstintegrals}
\[
q_\mu U^\mu = 0 \Rightarrow (\nabla_U q_\mu) U^\mu + q_\mu (\nabla_U U^\mu) = 0 \Leftrightarrow q_\mu P^\mu = 0.
\]
Differentiating this new first integral yields
\begin{equation} \label{relation}
(\nabla_U q_\mu) P^\mu + q_\mu (\nabla_U P^\mu) = 0 \Leftrightarrow q_\mu q^\mu = - K,
\end{equation}
where 
\[
K = - R_{\alpha\beta\mu\nu}P^\alpha U^\beta P^\mu U^\nu
\]
is the sectional curvature of the tangent plane spanned by $U$ and $P$. The magnitude $a(\tau)$ of the proper acceleration must then be chosen so that \eqref{relation} holds at each instant. Using \eqref{diffeqns} and \eqref{firstintegrals}, we can easily rewrite \eqref{relation} as
\begin{equation} \label{relation2}
(\nabla_U P_\mu) (\nabla_U P^\mu) + a^2 = - K.
\end{equation}
%
%
\section{The G\"odel universe}\label{section2}
The G\"odel universe is a space-time homogeneous solution of the Einstein equations with a pressureless perfect fluid and a negative cosmological constant. It corresponds to the Lorentzian metric in $\bbR^4$ given in cylindrical coordinates by
\[
ds^2 = - \left(dt + \sqrt{2}(1-\cosh r) d\varphi\right)^2 + dr^2 + \sinh^2 r \, d\varphi^2 + dz^2,
\]
where we have chosen units\footnote{Notice that none of adimensional quantities which we will compute -- e.g.~angles and velocities, including the total integrated acceleration -- depend on the choice of units.} such that the cosmological constant is $\Lambda = - \frac12$ and the matter density is $\rho = \frac{1}{8\pi}$ (see for instance\footnote{There is a sign difference corresponding to a different choice of the direction of rotation of the fluid elements, which we take to be anti-clockwise.} \cite{OS03}). The fluid worldlines are tangent to the timelike Killing vector field $\frac{\partial}{\partial t}$, and have vorticity given in our units by $\omega = \frac{\sqrt{2}}2 \frac{\partial}{\partial z}$ (so the fluid elements are rotating with angular velocity $\frac{\sqrt{2}}2$ about the $z$-axis with respect to the local compass of inertia). 

In what follows we will restrict ourselves to a totally geodesic submanifold of constant $z$ (which we will still call the G\"odel universe), with metric
\[
ds^2 = - \left(dt + \sqrt{2}(1-\cosh r) d\varphi\right)^2 + dr^2 + \sinh^2 r \, d\varphi^2
\]
(the nontrivial part of the G\"odel metric). Notice that the quotient of this manifold by the isometric $\bbR$-action generated by $\frac{\partial}{\partial t}$ is the hyperbolic plane of radius $1$ (hence our choice of units).

The G\"odel universe famously contains closed timelike curves (for instance the integral curves of $\frac{\partial}{\partial \varphi}$ for $r$ sufficiently large). For any timelike curve with proper time $\tau$ we have
\[
-d\tau^2 = - \left(dt + \sqrt{2}(1-\cosh r) d\varphi\right)^2 + dl^2,
\]
where $dl^2$ is the line element of the hyperbolic plane. Solving for $dt$ yields
\[
dt = \sqrt{2}(\cosh r-1) d\varphi + \sqrt{dl^2+d\tau^2}
\]
for future-directed curves (corresponding to the positive square root). Therefore if the projection $\gamma$ of the timelike curve on the hyperbolic plane is closed then the coordinate time difference between its endpoints is
\[
\Delta t = \oint_\gamma dt = \sqrt{2} \oint_\gamma (\cosh r-1) d\varphi + \oint_\gamma \sqrt{1+\left(\frac{d\tau}{dl}\right)^2} dl.
\]
It is easy to check that the velocity $v$ of the timelike curve with respect to the matter satisfies
\[
\frac{dl}{d\tau}=\frac{v}{\sqrt{1-v^2}} \Rightarrow  \sqrt{1+\left(\frac{d\tau}{dl}\right)^2} = \frac1{v}.
\]
If the curve $\gamma$ is simple and bounds a region $D$ then we can use the Stokes theorem to write
\begin{equation} \label{Delta t}
\Delta t = \sqrt{2} \iint_D \sinh r \, dr \wedge d\varphi + \oint_\gamma \frac{dl}{v} = \sqrt{2} A + \oint_\gamma \frac{dl}{v},
\end{equation}
where $A$ is the oriented hyperbolic area of $D$ (negative or positive according to whether $\gamma$ is traversed clockwise or anticlockwise). For the timelike curve to be closed ($\Delta t = 0$) we then see that $\gamma$ must be traversed clockwise (so that $A<0$). Moreover, since we must necessarily have $v<1$, the length $l$ of $\gamma$ satisfies
\begin{equation} \label{ineq}
l < \sqrt{2} |A|.
\end{equation}
Given $|A|$, we known that $l$ is minimized when $\gamma$ is a circle \cite{Osserman78, HHM99}. Now a circle of radius $r$ in the hyperbolic plane satisfies
\[
l = 2 \pi \sinh r \qquad \text{ and } \qquad |A| = 2 \pi (\cosh r - 1),
\]
and so \eqref{ineq} is satisfied by a circle of radius $r$ if
\[
\sinh r < \sqrt{2} (\cosh r - 1),
\]
that is, if $r$ is bigger than the positive root of
\[
\sinh^2 r = 2(\cosh r - 1)^2 \Leftrightarrow \cosh^2 r - 4 \cosh r + 3 = 0,
\]
which is $r=\arccosh 3$. We conclude that \eqref{ineq} can only hold if $|A| > 4\pi$. In other words, the projection $\gamma$ of a closed timelike curve must enclose a sufficiently large area. Notice that if $|A| > 4\pi$ then there is indeed a closed timelike curve whose projection encloses this area -- just choose $\gamma$ to be a circle and $v<1$ an appropriate constant. Also, the isoperimetric property of the circle implies that the the projection of a closed timelike curve must have length $l > 2\pi\sinh(\arccosh(3))=4\pi\sqrt{2}$. Moreover, we see from \eqref{Delta t} that the maximum velocity of the closed timelike curve with respect to the matter, $v_{max}$, must satisfy
\[
\sqrt{2} A + \frac{l}{v_{max}} \leq 0 \Leftrightarrow v_{max} \geq \frac{l}{\sqrt{2}|A|} \geq \frac{\sinh r}{\sqrt{2}(\cosh r - 1)} > \frac{\sqrt{2}}{2},
\]
again by the isoperimetric property of the circle (an equivalent observation was made in \cite{CGL83}).

Choosing the natural orthonormal frame
\[
\begin{cases}
E_0 = \frac{\partial}{\partial t} \\ 
E_r = \frac{\partial}{\partial r} \\ 
E_\varphi = \frac1{\sinh r}\frac{\partial}{\partial \varphi} + \frac{\sqrt{2}(\cosh r - 1)}{\sinh r} \frac{\partial}{\partial t} 
\end{cases}
\]
with dual coframe
\[
\begin{cases}
\omega^0 = dt + \sqrt{2}(1-\cosh r) d\varphi \\
\omega^r = dr \\
\omega^\varphi = \sinh r \, d\varphi
\end{cases}
\]
one readily obtains from the first Cartan structure equations
\[
d \omega^\alpha + \omega^\alpha_{\,\,\,\beta} \wedge \omega^\beta = 0
\]
the nonvanishing connection forms
\begin{equation}\label{connection}
\begin{cases}
\omega^0_{\,\,\,r} = \omega^r_{\,\,\,0} = -\frac{\sqrt{2}}{2} \omega^\varphi \\
\omega^0_{\,\,\,\varphi} = \omega^\varphi_{\,\,\,0} = \frac{\sqrt{2}}{2} \omega^r \\
\omega^r_{\,\,\,\varphi} = - \omega^\varphi_{\,\,\,r} = - \frac{\sqrt{2}}{2} \omega^0 - \cotanh r \, \omega^\varphi
\end{cases}
\end{equation}
The second Cartan structure equations
\[
\Omega^\alpha_{\,\,\,\beta} = d \omega^\alpha_{\,\,\,\beta} + \omega^\alpha_{\,\,\,\gamma} \wedge  \omega^\gamma_{\,\,\,\beta}
\]
then yield the nonvanishing curvature forms
\[
\begin{cases}
\Omega^0_{\,\,\,r} = \Omega^r_{\,\,\,0} = - \frac12 \omega^0 \wedge \omega^r \\
\Omega^0_{\,\,\,\varphi} = \Omega^\varphi_{\,\,\,0} = - \frac12 \omega^0 \wedge \omega^\varphi \\
\Omega^r_{\,\,\,\varphi} = - \Omega^\varphi_{\,\,\,r} = \frac12 \omega^r \wedge \omega^\varphi
\end{cases}
\]
From the general expression
\[
\Omega^\alpha_{\,\,\,\beta} = \frac12 R^\alpha_{\,\,\,\beta\mu\nu} \, \omega^\mu \wedge \omega^\nu
\]
we obtain the independent components of the Riemann tensor in this orthonormal frame,
\begin{equation} \label{curvature}
R_{0r0r} = R_{0 \varphi 0 \varphi} = R_{r \varphi r \varphi} = \frac12.
\end{equation}

Consider now a timelike curve whose projection on the hyperbolic plane is a circle of radius $r$. Because of the space-time homogeneity of the G\"odel universe, we may assume that the circle is centered at the origin $r=0$ of our coordinate system. If the circle is traversed clockwise with constant velocity $v=\tanh u$ with respect to the matter then the four-velocity of the timelike curve is
\[
U = \cosh u \, E_0 - \sinh u \, E_\varphi,
\]
and hence its covariant acceleration is
\begin{align}
\nabla_U U & = \cosh u \, \omega^\alpha_{\,\,\,0}(U)  E_\alpha - \sinh u \, \omega^\alpha_{\,\,\,\varphi}(U) E_\alpha \label{covarianta} \\
& = \sinh u (\sqrt{2} \cosh u - \sinh u \cotanh r) E_r. \nonumber
\end{align}
In particular, the curve is a geodesic if and only if $v = \sqrt{2} \tanh r$. The space-time homogeneity of the G\"odel universe implies that all timelike geodesics of the G\"odel universe are of this form. Notice that the radius of the circle satisfies
\[
\tanh r < \frac{\sqrt{2}}{2} \Leftrightarrow r < \arccosh(\sqrt{2}),
\]
and so no timelike geodesic can be closed.
%
%
\section{The Jacobi equation}\label{section3}
The minimum conditions \eqref{diffeqns}, \eqref{firstintegrals} on a free-fall arc reduce to the Jacobi equation for a Jacobi field orthogonal to the four-velocity. If we choose coordinates $(r,\varphi)$ centered at the center of the free-fall arc, we have
\[
U = \cosh u \, E_0 - \sinh u \, E_\varphi
\] 
with $\tanh u = \sqrt{2} \tanh r$, and so
\[
P = P^r E_r + \widetilde{P}^\varphi \widetilde{E}_\varphi,
\]
where
\[
\widetilde{E}_\varphi = - \sinh u \, E_0 + \cosh u \, E_\varphi
\]
is the unit vector field orthogonal to $U$ and $E_r$ pointing in the same direction as $E_\varphi$. Using the connection forms \eqref{connection} and $\tanh u = \sqrt{2} \tanh r$, one readily checks that
\[
\begin{cases}
\nabla_U E_r = - \frac{\sqrt{2}}2 \widetilde{E}_\varphi \\
\nabla_U \widetilde{E}_\varphi = \frac{\sqrt{2}}2 E_r
\end{cases}
\]
for a geodesic, and so
\[
\nabla_U P = \left( \dot{P}^r + \frac{\sqrt{2}}2 \widetilde{P}^\varphi \right) E_r + \left( \dot{\widetilde{P}}\,\!^\varphi - \frac{\sqrt{2}}2 P^r \right) \widetilde{E}_\varphi,
\]
whence
\[
\nabla_U\nabla_U P = \left( \ddot{P}^r + \sqrt{2} \dot{\widetilde{P}}\,\!^\varphi - \frac12 P^r \right) E_r + \left( \ddot{\widetilde{P}}\,\!^\varphi - \sqrt{2} \dot{P}^r - \frac12 \widetilde{P}^\varphi \right) \widetilde{E}_\varphi.
\]
The Jacobi equation,
\[
\nabla_U \nabla_U P_\mu = - R_{\mu\alpha\beta\gamma} U^\alpha P^\beta U^\gamma,
\]
 has three components. The component along $U$,
\[
(\nabla_U \nabla_U P_\mu) U^\mu = - R_{\mu\alpha\beta\gamma} U^\mu U^\alpha P^\beta U^\gamma,
\]
is trivially satisfied. The component along $E_r$,
\[
\ddot{P}^r + \sqrt{2} \dot{\widetilde{P}}\,\!^\varphi - \frac12 P^r = - R_{r0r0} \cosh^2 u \, P^r - R_{r\varphi r\varphi} \sinh^2 u \, P^r,
\]
and the component along $E_\varphi$,
\[
\cosh u \left(\ddot{\widetilde{P}}\,\!^\varphi - \sqrt{2} \dot{P}^r - \frac12 \widetilde{P}^\varphi\right) = - R_{\varphi 0 \varphi 0} \cosh^2 u \, P^\varphi + R_{\varphi 00\varphi} \cosh u \, \sinh u \, P^0,
\]
form the system of linear ODEs
\[
\begin{cases}
\ddot{P}^r + \sqrt{2} \dot{\widetilde{P}}\,\!^\varphi = - \sinh^2 u \, P^r \\
\ddot{\widetilde{P}}\,\!^\varphi - \sqrt{2} \dot{P}^r = 0
\end{cases}
\]
(where we used \eqref{curvature}). This system is readily solved to yield
\[
\begin{cases}
P^r = - \frac{\sqrt{2}A}{\omega^2} + B \cos(\omega \tau) - C \sin(\omega \tau) \\
\widetilde{P}^\varphi = A \left( 1 - \frac2{\omega^2} \right) \tau + \frac{\sqrt{2}B}{\omega} \sin(\omega \tau) + \frac{\sqrt{2}C}{\omega} \cos(\omega \tau) + D
\end{cases}
\]
where $A,B,C,D \in \bbR$ are integrations constants and $\omega = \sqrt{\cosh^2 u + 1}$. Notice that the period of the projection of the timelike geodesic on the hyperbolic plane is precisely
\[
\frac{2\pi\sinh r}{\sinh u} = \frac{2\pi\tanh r}{\sinh u \sqrt{1-\tanh^2 r}} = \frac{\sqrt{2} \pi\tanh u}{\sinh u \sqrt{1- \frac12 \tanh^2 u}} = \frac{2\pi}{\omega},
\]
that is, $\omega$ is the angular frequency with respect to the proper time $\tau$.
%
%
\section{Optimal circular accelerated arcs}\label{section4}
For an optimal circular accelerated arc the covariant acceleration \eqref{covarianta} is proportional to $E_r$, and so $P = \pm E_r$. Using the connection forms \eqref{connection} one readily checks that
\[
\nabla_U E_r = \frac{\sqrt{2}}2 \sinh u \, E_0 + \left(\frac{\sqrt{2}}2 \cosh u - \cotanh r \sinh u \right) E_\varphi,
\]
and so
\[
(\nabla_U P_\mu) (\nabla_U P^\mu) = \frac12 - \sqrt{2} \sinh u \cosh u \cotanh r + \cotanh^2 r \sinh^2 u.
\]
From \eqref{covarianta} we have
\[
a^2 = \sinh^2 u \left(2 \cosh^2 u - 2 \sqrt{2} \sinh u \cosh u \cotanh r + \sinh^2 u \cotanh^2 r\right),
\]
and from \eqref{curvature}
\[
K = - R_{\alpha\beta\mu\nu}P^\alpha U^\beta P^\mu U^\nu = - \cosh^2 u R_{r0r0} - \sinh^2 u R_{r\varphi r\varphi} = -\frac12\cosh^2 u -\frac12\sinh^2 u.
\]
Therefore we can write \eqref{relation2} as
\[
\tanh^2 r - \frac{\sqrt{2}}{v} \tanh r + \frac1{1+v^2} = 0,
\]
where $v=\tanh u$ is the constant velocity of the optimal accelerated arc with respect to the matter. This can be solved with respect to $\tanh r$ to yield
\begin{equation} \label{branches}
\tanh r = \frac1{\sqrt{2}v}\left(1\pm\sqrt{\frac{1-v^2}{1+v^2}}\right)
\end{equation}
The $\pm$ sign determines whether $\nabla_U U$ points in the same direction as $E_r$ or in the opposite direction. Indeed, from \eqref{covarianta} we see that this happens if and only if
\[
\sqrt{2}-\frac{v}{\tanh r} > 0 \Leftrightarrow \sqrt{2} v\tanh r > v^2 \Leftrightarrow 1\pm\sqrt{\frac{1-v^2}{1+v^2}} > v^2 \Leftrightarrow \pm \frac{1}{\sqrt{1-v^4}} > -1.
\]
It is easy to see that as $v$ varies from $0$ to $1$ the $-$ branch of \eqref{branches} increases from $0$ to $\frac{\sqrt{2}}{2}$, whereas the $+$ branch decreases from $+ \infty$ to $\frac{\sqrt{2}}{2}$. Since we must have $\tanh r<1$, the $+$ branch is only defined for $v>v_0\simeq 0.916126$.

Since $\tanh(\arccosh(3))=\frac{2\sqrt{2}}{3}>\frac{\sqrt{2}}{2}$, we see that only the $+$ branch of \eqref{branches} can contain an optimal circular accelerated arc which is a closed timelike curve. Such a curve must satisfy
\[
\sqrt{2} A + \oint_\gamma \frac{dl}{v} = 0 \Leftrightarrow -2\pi\sqrt{2}(\cosh r - 1) + \frac{2\pi}{v} \sinh r = 0 \Leftrightarrow v = \frac{\sinh r}{\sqrt{2} (\cosh r - 1)}.
\]
Therefore we must have
\[
\frac{1-v^2}{1+v^2} = \frac{2(\cosh r - 1)^2-\sinh^2 r}{2(\cosh r - 1)^2+\sinh^2 r} = \frac{\cosh r - 3}{3\cosh r - 1}
\]
and so from \eqref{branches}
\[
\frac{\sinh^2 r}{\cosh r (\cosh r - 1)} = 1 + \sqrt{\frac{\cosh r - 3}{3\cosh r - 1}} \Leftrightarrow \frac1{\cosh r} = \sqrt{\frac{\cosh r - 3}{3\cosh r - 1}}.
\]
Squaring this equation we obtain
\begin{equation} \label{optimalaccelerated}
\cosh^3 r - 3 \cosh^2 r - 3 \cosh r + 1 = 0,
\end{equation}
and, since $x=-1$ is an obvious root of the polynomial $x^3-3x^2-3x+1$, we readily obtain $r=\arccosh(2+\sqrt{3})$ as the only real solution of \eqref{optimalaccelerated}.

This optimal closed accelerated arc is precisely Malament's conjectured optimal trajectory. Its total integrated acceleration is
\[
TA = a \frac{2 \pi \sinh r}{\sinh u} = \sinh u (\sqrt{2} \cosh u - \sinh u \cotanh r) \frac{2 \pi \sinh r}{\sinh u},
\]
or, since
\[
\tanh u = \frac{\sinh r}{\sqrt{2} (\cosh r - 1)} \Leftrightarrow \cosh u \sinh r = \sqrt{2} \sinh u (\cosh r - 1),
\]
we have
\[
TA = 2 \pi \sinh u (\cosh r - 2).
\]
Using
\[
\begin{cases}
\cosh r = 2+\sqrt{3} \\
\sinh r = \sqrt{\cosh^2r - 1} = \sqrt{6 + 4\sqrt{3}} \\
v= \frac{\sinh r}{\sqrt{2} (\cosh r - 1)} = \sqrt{\frac{\sqrt{3}}{2}} \\
\cosh u = \frac{1}{\sqrt{1-v^2}} = \sqrt{4+2\sqrt{3}} \\
\sinh u = \frac{v}{\sqrt{1-v^2}} = \sqrt{3+2\sqrt{3}}
\end{cases}
\]
we obtain
\[
TA = 2\pi\sqrt{9+6\sqrt{3}} \simeq 27.6691,
\]
as advertised in the introduction.

The optimality of this curve can be partially checked as follows. Using the formulae above, it is not hard to show that
\[
TA = 2\pi(\cosh r - 2)\sqrt{\frac{\cosh r + 1}{\cosh r - 3}}
\]
for closed timelike curves with constant velocity with respect to the matter whose projection on the hyperbolic plane is a circle of radius $r$. It is then easy to show that Malament's curve does indeed minimize $TA$ among these curves.

Notice that the proper time necessary to traverse Malament's curve is
\[
\frac{2 \pi \sinh r}{\sinh u} = 2 \sqrt{2} \pi,
\]
which in our units can be written as
\[
\frac{2 \sqrt{2} \pi}{\sqrt{8 \pi \rho}} = \sqrt{\frac{\pi}{\rho}}.
\]
Therefore if the matter density of the G\"odel universe is comparable to the matter density of our universe then the proper time for this curve is comparable to the Hubble time for our universe, about $10^{10}$ years. The proper acceleration will be correspondingly low.
%
%
\section{The optimal trajectory}\label{section5}
\begin{figure}[h!]
\begin{center}
\psfrag{a}{$\alpha$}
\psfrag{b}{$\beta$}
\psfrag{e}{$\varepsilon$}
\psfrag{f}{$\varphi$}
\psfrag{p}{$\psi$}
\psfrag{t}{$\theta$}
\psfrag{R}{$R$}
\psfrag{r}{$r$}
\psfrag{x}{$x$}
\psfrag{y}{$y$}
\epsfxsize=.7\textwidth
\leavevmode
\epsfbox{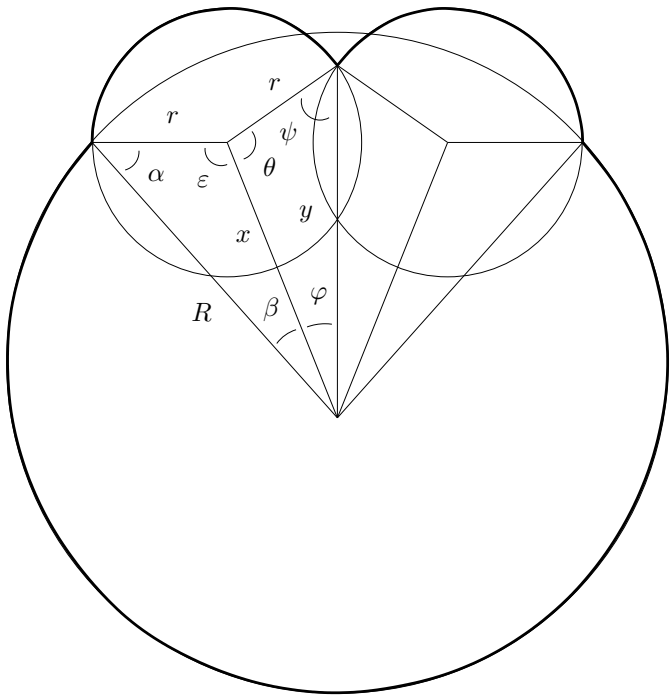}
\end{center}
\caption{Projection of the optimal trajectory on the hyperbolic plane (circles and lines represent circles and geodesics of the hyperbolic plane).}\label{Optimal_Godel}
\end{figure}

We now present our candidate for the optimal trajectory and give evidence for its optimality. Our simplifying assumption is that this trajectory contains a circular accelerated arc of radius $R>0$. Assuming, by analogy with Malament's trajectory, that the acceleration points outwards, we have from \eqref{branches} that the velocity $V$ of this arc with respect to the matter satisfies
\begin{equation} \label{R}
\tanh R = \frac1{\sqrt{2}V}\left(1+\sqrt{\frac{1-V^2}{1+V^2}}\right).
\end{equation}
Because we are simply looking for the closed timelike curve which minimizes the total integrated acceleration, we do not specify the initial and final four-velocities, and so, as stated in section~\ref{section1}, $P$ must vanish at the endpoints. Since $\rho=1$ on accelerated arcs and instantaneous accelerations, we conclude that these endpoints must be on free-fall arcs. Moreover, we have $P=E_r$ throughout the accelerated arc, and so if there are instantaneous accelerations at its endpoints these must be in the outward radial direction. The simplest candidate curve of this kind is an initial free-fall arc, followed by the circular accelerated arc, followed by another free-fall arc, possibly with instantaneous accelerations at the transitions. By simplicity we assume that the final free-fall arc mirrors the initial one, both having radius $r>0$ and velocity $v=\sqrt{2} \tanh r$ with respect to the matter. The projection of this candidate curve on the hyperbolic plane will then be an arc of a large circle of radius $R$, and two arcs of smaller circles of radius $r$, as shown in Figure~\ref{Optimal_Godel} (where we use circles and lines to represent circles and geodesics of the hyperbolic plane).

Each trajectory of this kind is determined by $3$ parameters, which we will take to be the velocity $V$ of the accelerated arc with respect to the matter, the boost parameter $\Delta u$ of the instantaneous accelerations, and the angle $\theta$ in Figure~\ref{Optimal_Godel}. This immediately fixes $R$ through \eqref{R}. On the other hand, if we use coordinates centered at the center of the accelerated arc then the four-velocity of this arc is
\[
U_{-} = \Gamma E_0 - V \Gamma E_\varphi,
\]
where
\[
\Gamma = \frac1{\sqrt{1-V^2}}.
\]
If we apply an instantaneous acceleration in the radial direction with boost parameter $\Delta u$ then the four-velocity jumps to
\begin{align*}
U_{+} & = \cosh (\Delta u) U_{-} + \sinh (\Delta u) E_r \\
& = \Gamma \cosh (\Delta u) E_0 + \sinh (\Delta u) E_r - V \Gamma \cosh (\Delta u) E_\varphi.
\end{align*}
On the other hand, we must also have
\[
U_{+} = \gamma E_0 + v \gamma \sin \alpha E_r - v \gamma \cos \alpha E_\varphi,
\]
where $v$ is the velocity of the final (hence also the initial) free-fall arc with respect to the matter,
\[
\gamma = \frac1{\sqrt{1-v^2}}
\]
and $\alpha$ is the angle between the two radii depicted in Figure~\ref{Optimal_Godel} (notice that the radius of a circle in the hyperbolic plane is orthogonal to the circle, and so $\alpha$ is also the angle between the two tangents). Therefore
\[
\gamma = \Gamma \cosh (\Delta u) \Leftrightarrow v = \frac{\sqrt{\sinh^2 (\Delta u) + V^2}}{\cosh (\Delta u)}.
\]
Notice that this fixes $r = \arctanh \left(\frac{\sqrt{2}v}{2}\right)$. Moreover, we have
\[
\tan \alpha = \frac{\tanh(\Delta u)}{V\Gamma},
\]
and so $\alpha$ is also fixed. Now if $x$ is the distance between the center of the accelerated arc and the center of the free-fall arc, we have from the hyperbolic law of cosines \cite{Anderson07}
\[
\cosh x = \cosh r \cosh R - \sinh r \sinh R \cos \alpha
\]
(see Figure~\ref{Optimal_Godel}), which fixes $x$. From the hyperbolic law of sines we have
\[
\sin \beta = \frac{\sinh r}{\sinh x} \sin \alpha
\]
and
\[
\sin \varepsilon = \frac{\sinh R}{\sinh x} \sin \alpha,
\]
which fixes the angles $\beta$ and $\varepsilon$ depicted in Figure~\ref{Optimal_Godel}. Similarly, if  $y$ is the distance between the center of the accelerated arc and the outermost intersection of the free-fall arcs, we have from the hyperbolic law of cosines
\[
\cosh y = \cosh r \cosh x - \sinh r \sinh x \cos \theta
\]
(see Figure~\ref{Optimal_Godel}), which fixes $y$. From the hyperbolic law of sines we have
\[
\sin \varphi = \frac{\sinh r}{\sinh y} \sin \theta
\]
and
\[
\sin \psi = \frac{\sinh x}{\sinh y} \sin \theta,
\]
which fixes the final two angles $\varphi$ and $\psi$ depicted in Figure~\ref{Optimal_Godel}.

The three parameters $V$, $\Delta u$ and $\theta$ are not independent, as we are looking for a closed timelike curve. Equation \eqref{Delta t} implies that the area $A$ enclosed by the projection curve in the hyperbolic plane must satisfy
\begin{equation} \label{closed}
A = \frac{L}{\sqrt{2}V} + \frac{l}{\sqrt{2}v},
\end{equation}
where
\begin{equation} \label{L}
L = 2 (\pi - \beta - \varphi) \sinh R
\end{equation}
is the length of the accelerated arc, and
\begin{equation} \label{l}
l = 2 (2\pi - \varepsilon - \theta) \sinh r
\end{equation}
is the sum of the lengths of the free-fall arcs. Using the Gauss-Bonnet formula for the area of a hyperbolic triangle \cite{Anderson07}, we can write
\begin{align} \label{A}
& A = 2 (\pi - \beta - \varphi) (\cosh R - 1) + 2 (2\pi - \varepsilon - \theta) (\cosh r - 1) \\
\nonumber & \qquad \qquad \qquad \qquad + 2 (\pi - \alpha - \beta - \varepsilon) + 2 (\pi - \varphi - \theta - \psi).
\end{align}

From \eqref{covarianta}, the acceleration the accelerated arc is
\[
a = V \Gamma^2 (\sqrt{2} - V \cotanh R),
\]
and the proper time necessary to traverse it is
\[
\frac{2(\pi-\beta-\varphi) \sinh R}{V \Gamma}.
\]
Therefore the trajectory's total integrated acceleration (including the two instantaneous accelerations) is
\begin{equation} \label{cost}
TA = 2 \Gamma (\pi-\beta-\varphi)(\sqrt{2} \sinh R - V \cosh R) + 2 \Delta u.
\end{equation}

Notice that the angle between the projections of the two free-fall arcs is $2\psi$. This means that if $\{F_0, F_r, F_\varphi\}$ is the orthonormal frame associated to coordinates $(r,\varphi)$ centered at the center of the final free-fall arc, so that the final four-velocity is given by
\[
U_1 = \gamma F_0 - v \gamma F_\varphi,
\]
then the initial four-velocity is
\[
U_0 = \gamma F_0 + v \gamma \sin(2\psi) F_r - v \gamma \cos(2\psi) F_\varphi.
\]
Therefore the hyperbolic angle $\Delta u'$ between the corresponding four-velocities satisfies
\[
- \cosh \Delta u' = \left\langle U_0, U_1 \right\rangle = - \gamma^2 +  v^2 \gamma^2 \cos(2\psi),
\]
where $\langle \cdot, \cdot \rangle$ is the metric, and so the integrated acceleration necessary to boost the final four-velocity to the initial four-velocity is
\[
\Delta u' = \arccosh \left(\frac{1 - v^2 \cos(2\psi)}{1 - v^2}\right).
\]
If one wishes to regard the candidate curve as periodic then one must add this value of $\Delta u'$ to \eqref{cost}.

We are now ready to search for the optimal curve. Because of the complexity of the formulae above, a numerical search seems to be inevitable. We use a shooting algorithm as follows: starting with arbitrary values for $\Delta u$, $\theta$ and $V$, we determine the remaining parameters $R,r,x,y,\alpha,\beta,\varepsilon,\varphi,\psi$. We then use \eqref{L}, \eqref{l} and \eqref{A} to compute $L$, $l$ and $A$, and check whether \eqref{closed} holds; if not, we modify the value of $V$ until it does. At this point we have a closed timelike curve and can use \eqref{cost} to compute its total integrated acceleration. Using this algorithm, we vary $\Delta u$ (keeping $\theta$ fixed) until $TA$ is minimized, and then vary $\theta$ (keeping $\Delta u$ fixed) until $TA$ is again minimized. Repeating these iterations the method quickly converges to yield a closed timelike curve whose total integrated acceleration is minimum among all curves of this type. The solution is determined by
\begin{equation} \label{mainparameters_numerical}
\begin{cases}
V \simeq 0.930597 \\
\Delta u \simeq 1.16765 \\
\theta \simeq 1.64430 \\
\end{cases}
\end{equation}
We notice in passing that the velocity of the accelerated arc is very close to, but slightly smaller than, the velocity of Malament's curve, which is
\[
\sqrt{\frac{\sqrt{3}}2} \simeq 0.930605.
\]
The remaining parameters of the optimal trajectory are
\begin{equation} \label{secondaryparameters_numerical}
\begin{cases}
v \simeq 0.978202 \\
r \simeq 0.851195 \\
R \simeq 1.99194 \\
\alpha \simeq 0.313261 \\
x \simeq 1.25222 \\
\beta \simeq 0.184815 \\
\varepsilon \simeq 2.37999 \\
y \simeq 1.66316 \\
\varphi \simeq 0.385023 \\
\psi \simeq 0.681328
\end{cases}
\end{equation}
yielding the total integrated acceleration
\[
TA \simeq 24.9927.
\]
If we boost the final four-velocity to the initial four-velocity to make the trajectory periodic then the total integrated acceleration becomes
\[
TA \simeq 28.6085,
\]
in agreement with Malament's conjecture.

We now present evidence that this is indeed the optimal solution we seek. To avoid confusion, we keep $\{E_0,E_r,E_\varphi\}$ as the orthonormal frame associated to coordinates $(r,\varphi)$ centered at the center of the accelerated arc, and let $\{F_0,F_r,F_\varphi\}$ be the orthonormal frame associated to coordinates $(r,\varphi)$ centered at the center of the final free-fall arc (notice that $E_0=F_0$). Recall from section~\ref{section3} that on the final free-fall arc the primer has components
\begin{equation}\label{Jacobi}
\begin{cases}
P^r = - \frac{\sqrt{2}A}{\omega^2} + B \cos(\omega \tau) - C \sin(\omega \tau) \\
\widetilde{P}^\varphi = A \left( 1 - \frac2{\omega^2} \right) \tau + \frac{\sqrt{2}B}{\omega} \sin(\omega \tau) + \frac{\sqrt{2}C}{\omega} \cos(\omega \tau) + D
\end{cases}
\end{equation}
in the orthonormal frame $\{U, F_r, \widetilde{F}_\varphi\}$, where
\[
U = \gamma F_0 - v \gamma F_\varphi, \qquad \widetilde{F}_\varphi = - v \gamma F_0 + \gamma F_\varphi,
\]
$A,B,C,D \in \bbR$ are integration constants and
\[
\omega = \sqrt{\gamma^2 + 1} = \sqrt{\frac{2-v^2}{1-v^2}}.
\]
If for simplicity we set $\tau=0$ as the instant of transition from the accelerated arc to the free-fall arc, we have
\begin{equation} \label{P_at_tau_zero}
\begin{cases}
P^r = - \frac{\sqrt{2}A}{\omega^2} + B \\
\widetilde{P}^\varphi = \frac{\sqrt{2}C}{\omega} + D
\end{cases}
\end{equation}
at that instant. Moreover, it is easy to check that
\begin{equation} \label{dot_P_at_tau_zero}
\begin{cases}
\dot{P}^r = - C \omega \\
\dot{\widetilde{P}}\,\!^\varphi = A \left( 1 - \frac2{\omega^2} \right) + \sqrt{2}B
\end{cases}
\end{equation}
and
\begin{equation} \label{ddot_P_at_tau_zero}
\begin{cases}
\ddot{P}^r = - B \omega^2 \\
\ddot{\widetilde{P}}\,\!^\varphi = - \sqrt{2} \omega C
\end{cases}
\end{equation}
at the same instant.

Now, as we saw above, the instantaneous acceleration boosts the four-velocity from
\[
U_{-} = \Gamma E_0 - V \Gamma E_\varphi,
\]
to
\[
U_{+} = \Gamma \cosh (\Delta u) E_0 + \sinh (\Delta u) E_r - V \Gamma \cosh (\Delta u) E_\varphi.
\]
Consequently the new tangential direction is given by (minus) the normalized spatial velocity,
\[
F_\varphi = \frac{- \sinh (\Delta u) E_r + V \Gamma \cosh (\Delta u) E_\varphi}{\sqrt{\sinh^2 (\Delta u) + V^2 \Gamma^2 \cosh^2 (\Delta u)}},
\]
and consequently the new radial direction is given by
\[
F_r = \frac{V \Gamma \cosh (\Delta u) E_r + \sinh (\Delta u) E_\varphi}{\sqrt{\sinh^2 (\Delta u) + V^2 \Gamma^2 \cosh^2 (\Delta u)}}.
\]
On the other hand, the primer is boosted from
\[
P_{-} = E_r
\]
to
\begin{align*}
P_{+} & = \sinh (\Delta u) U_{-} + \cosh (\Delta u) E_r \\
& = \Gamma \sinh (\Delta u) E_0 + \cosh (\Delta u) E_r - V \Gamma \sinh (\Delta u) E_\varphi.
\end{align*}
Therefore the components of the primer at the beginning of the free-fall arc are
\begin{equation} \label{P^r}
P^r = \left\langle P_{+}, F_r \right\rangle = \frac{V\Gamma}{\sqrt{\sinh^2 (\Delta u) + V^2 \Gamma^2 \cosh^2 (\Delta u)}}
\end{equation}
and
\begin{equation} \label{P^varhi}
\widetilde{P}^\varphi = \left\langle P_{+}, \widetilde{F}_\varphi \right\rangle = \frac1{\gamma} \left\langle P_{+}, F_\varphi \right\rangle = \frac{- \Gamma \sinh(\Delta u)}{\sqrt{\sinh^2 (\Delta u) + V^2 \Gamma^2 \cosh^2 (\Delta u)}},
\end{equation}
where we used
\[
F_\varphi = v \gamma U_{+} + \gamma \widetilde{F}_\varphi \Leftrightarrow \widetilde{F}_\varphi = \frac1{\gamma} F_\varphi - v U_{+}
\]
and $\gamma = \Gamma \cosh (\Delta u)$.

Recall that we must have $P = 0$ at the endpoint of the final free-fall arc. Now from Figure~\ref{Optimal_Godel} it is clear that $\omega \tau = 2\pi - \varepsilon - \theta$ at this endpoint. From  \eqref{Jacobi}, \eqref{P_at_tau_zero}, \eqref{P^r} and \eqref{P^varhi} we then have
\[
\begin{cases}
- \frac{\sqrt{2}}{\omega^2} A + B = \frac{V\Gamma}{\sqrt{\sinh^2 (\Delta u) + V^2 \Gamma^2 \cosh^2 (\Delta u)}} \\
\frac{\sqrt{2}}{\omega}C + D = \frac{- \Gamma \sinh(\Delta u)}{\sqrt{\sinh^2 (\Delta u) + V^2 \Gamma^2 \cosh^2 (\Delta u)}} \\
- \frac{\sqrt{2}}{\omega^2}A + \cos(2\pi - \varepsilon - \theta) B - \sin(2\pi - \varepsilon - \theta) C = 0\\
\left( 1 - \frac2{\omega^2} \right) \frac{2\pi - \varepsilon - \theta}{\omega} A + \frac{\sqrt{2}}{\omega} \sin(2\pi - \varepsilon - \theta) B + \frac{\sqrt{2}}{\omega} \cos(2\pi - \varepsilon - \theta) C + D = 0
\end{cases}
\]
This can be regarded as a linear system in the integration constants $A,B,C,D \in \bbR$ which determine the primer in the final free-fall arc. Using the numerical data in \eqref{mainparameters_numerical} and \eqref{secondaryparameters_numerical} we obtain
\begin{equation} \label{ABCD}
\begin{cases}
A \simeq 1.04251 \\
B \simeq 0.600629 \\
C \simeq -0.572707 \\
D \simeq -0.677195
\end{cases}
\end{equation}
Now on the accelerated arc $\rho$ is identically $1$. Since, as stated in section~\ref{section1}, $\rho$ is a function of class $C^2$, we must have $\dot{\rho}=\ddot{\rho}=0$ at the instantaneous acceleration. Given that
\[
\rho^2 = (P^r)^2 + (\widetilde{P}^\varphi)^2,
\]
we have
\[
\rho \dot{\rho} = P^r \dot{P}^r + \widetilde{P}^\varphi \dot{\widetilde{P}}\,\!^\varphi
\]
and
\[
\dot{\rho}^2 + \rho \ddot{\rho} = \left(\dot{P}^r\right)^2 + \left(\dot{\widetilde{P}}\,\!^\varphi\right)^2 + P^r \ddot{P}^r + \widetilde{P}^\varphi \ddot{\widetilde{P}}\,\!^\varphi.
\]
Using \eqref{P_at_tau_zero}, \eqref{dot_P_at_tau_zero}, \eqref{ddot_P_at_tau_zero} and the numerical results in \eqref{ABCD} we then obtain
\[
\rho \dot{\rho} \simeq -8.71853 \times 10^{-6}
\]
and
\[
\dot{\rho}^2 + \rho \ddot{\rho} \simeq 2.79359 \times 10^{-5},
\]
at the instantaneous acceleration, which is very close to zero indeed. Notice that although $\rho=1$ at the instantaneous acceleration by construction, the values of $\dot{\rho}$ and $\ddot{\rho}$ on the final free-fall arc arise from the Jacobi equation, and are in principle completely independent. The fact that they both essentially vanish at the instantaneous acceleration is powerful evidence in favor of the optimality of our candidate curve.
%
%
\section{Conclusion}\label{section6}
In this paper we presented a candidate for the closed timelike curve in the G\"odel universe which minimizes the total integrated acceleration. The projection of the candidate curve on the hyperbolic plane representing the spatial part of the G\"odel universe is depicted in Figure~\ref{Optimal_Godel}, and is traversed clockwise. It starts with a free-fall arc of velocity $v \simeq 0.978202$ with respect to the matter, whose projection on the hyperbolic plane a $129^\circ$ arc of a circle of radius $r \simeq 0.851195$ in our units. This free-fall arc ends with an instantaneous acceleration of total integrated acceleration $\Delta u \simeq 1.16765$, which, as seen by the matter, rotates the spatial velocity anti-clockwise by $18^\circ$ and reduces its magnitude to $V \simeq 0.930597$. The candidate curve then coincides with a circular accelerated arc of constant velocity with respect to the matter, whose projection on the hyperbolic plane is a $295^\circ$ arc of a circle radius $R \simeq 1.99194$. This sequence of events is reversed at the end of the accelerated arc: there is a new instantaneous acceleration of total integrated acceleration $\Delta u \simeq 1.16765$, which, as seen by the matter, rotates the spatial velocity anti-clockwise by $18^\circ$ and increases its magnitude back to $v \simeq 0.978202$, after which there is a new free-fall arc whose projection on the hyperbolic plane is another $129^\circ$ arc of a circle of radius $r \simeq 0.851195$.

Notice that while our candidate curve does satisfy the minimum conditions, it is possible that it is just a local minimum, or even a saddle point. The question of whether it is indeed the minimum total integrated acceleration closed timelike curve in the G\"odel universe will have to be settled by different methods.
%
%

\end{document}